\documentclass[preprint,12pt]{elsarticle}
\usepackage{amssymb}

\journal{Frontiers in Computational Neuroscience }

\begin{document}

\begin{frontmatter}

\title{Metabolic efficiency with fast spiking in the squid axon}

\author{Abdelmalik Moujahid and Alicia d'Anjou}
\address{Computational Intelligence Group, Department of Computer Science, University of the Basque Country UPV/EHU, 20018 San Sebastian, Spain}

\date{November 15, 2012}

\begin{abstract}
Fundamentally, action potentials in the squid axon are consequence of the entrance of sodium ions during the depolarization of the rising phase of the spike mediated by the outflow of potassium ions during the hyperpolarization of the falling phase. Perfect metabolic efficiency with a minimum charge needed for the change in voltage during the action potential would confine sodium entry to the rising phase and potassium efflux to the falling phase. However, because sodium channels remain open to a significant extent during the falling phase, a certain overlap of inward and outward currents is observed. In this work we investigate the impact of ion overlap on the number of the adenosine triphosphate (ATP) molecules and energy cost required per action potential as a function of the temperature in a Hodgkin-Huxley model. Based on a recent approach to computing the energy cost of neuronal AP generation not based on ion counting, we show that increased firing frequencies induced by higher temperatures imply more efficient use of sodium entry, and then a decrease in the metabolic energy cost required to restore the concentration gradients after an action potential. Also, we determine values of sodium conductance at which the hydrolysis efficiency presents a clear minimum.
\end{abstract}



\end{frontmatter}

\maketitle

\section{Introduction}
\label{sec:1}

The generation of action potentials in mammalian neurons involves the flux of different ions such as sodium, potassium and calcium across the cell membrane. In this process, the electrochemical gradients are partially altered and must be restored by ion pumps which move ions from one side of the membrane to the other at the expense of energy. Reestablishing the concentration gradients after electrical discharges demands most of the energy used for neuronal metabolism \cite{Laughlin1998,Laughlin2001,Shulman2004,Siekevitz2004}. This requirement for metabolic energy has important implications for the brain's evolution and function \cite{Attwell2001}, and the availability of energy may impose a limit on neural activity taking into account that the brain has very small energy reserves \cite{Ames2000}. It is, however, generally accepted that energy metabolism is highly organized within cells resulting in energetically efficient mechanisms that transfer energy from the site of generation to the processes that require it \cite{Ames2000,Belanger2011}.

On other hand, all energy used for neural metabolism is finally transformed into heat \cite{Ames2000}, and the metabolic brain activation appears to be the primary cause of heat production. Because neural properties are temperature dependent, potential imbalance between heat production and dissipation could lead to overheating and aberrant functioning \cite{Koch1998,Falk1990,Kiyatkin2007}. Studying the relationship between temperature, firing frequency, sodium entry and the energy cost required to generate an action potential using neuron models like the Hodgkin-Huxley model  \cite{Hodgkin1975} will provides a useful framework for addressing these issues.

The Hodgkin-Huxley model representing the dynamics of the squid giant axon continues to be the most frequently used model to study the dynamics and other properties of actual neurons.
Based on biophysical considerations about the nature of the Hodgkin-Huxley model, we have recently found an analytical expression of the electrochemical energy involved in the dynamics of the model, which provides a new approach for estimating the energy consumption during the resting and active states of neurons  \cite{Moujahid2011}. This energy function was used as a measure to evaluate the metabolic energy consumption of a neuron to maintain its signalling activity and to estimate the metabolic cost of transmitting information between neurons.

This approach, contrary to other methods \cite{Attwell2001,Lennie2003}, does not require ion counting for estimating the metabolic energy consumption of the generation of action potentials, and gives us the opportunity to check in the Hodgkin-Huxley model which ion counting gives the correct metabolic energy consumption. In this work we investigate the impact of ion currents overlapping on the number of ATP molecules required to restore the concentration gradients after an action potential in the Hodgkin-Huxley model. Because the observed overlap is temperature dependent, we have computed the number of ATP molecules per action potential and its corresponding energy cost at different values of temperature.
Both the classic study by Hodgkin and Huxley of the squid axon \cite{Hodgkin1975}, and other recent works \cite{Attwell2001,Lennie2003} assume that the action potential requires four times Na$^+$ charge compared to the charge needed for the change in voltage.
This waste of Na$^+$ charge, and accordingly metabolic energy, is the result of extensive overlap between inward Na$^+$ and outward K$^+$ during the generation of action potentials.

However, it has been demonstrated that mammalian central neurons, characterized by action potentials similar to those of the squid giant axon, are significantly more efficient in generating action potentials \cite{Carter2009}.

We show in this work that increased firing frequencies induced by higher temperatures in the Hodgkin and Huxley model imply more efficient use of sodium entry and metabolic energy. The paper is organized as follows. In section 2, the dynamics and electrochemical energy of the Hodgkin and Huxley model are introduced. In section 3 we discuss the overlap of ion currents and energy efficiency as a function of temperature in the squid axon. Finally, conclusions are drawn in section 4.

\section{Materials and Methods}
\subsection{The Hodgkin-Huxley neuron energy}
\label{sec:2}

In the original Hodgkin-Huxley model \cite{Hodgkin1952}, the dynamics governing the membrane potential is given by:
\begin{equation}
\begin{array}{ll}
C \, \dot{V}= \small -g_{N{a}}m^{3}h(V-E_{N{a}})\\-g_{K}n^{4}(V-E_{K})-g_{l}(V-E_{l})+ I,\\
\end{array}
\label{equ1}
\end{equation}
where $V$ is the membrane potential in mV, $C$ the membrane capacitance density in $\mu$F/cm$^2$, $I$ is the total membrane current density in $\mu \textrm{A/cm}^2$. $g_{N_{a}}$, $g_{K}$ and $g_{l}$ are the maximal conductances per unit area for ion and leakage channels, and $E_{N{a}}$, $E_{K}$ and $E_{l}$ are the corresponding reversal potentials.

The gating variables $m$,$h$ and $n$, representing respectively sodium channels activation and deactivation variables, and potassium channels activation variable, obey the standard kinetic equation $\dot{x}=\alpha_x(1-x)-\beta_x x$, ($x=m,h,n$), where $\alpha_x$ and $\beta_x$ are voltage-dependent variables. For sodium channels, the activation and deactivation rates are given by,
\begin{eqnarray*}
\alpha_{m}(V)&=&(2.5-0.1V)/(\textrm{exp}\left(2.5-0.1V\right)-1),\\
\beta_{m}(V)&=&4\,\textrm{exp}\left(-V/18\right),\\
\alpha_{h}(V)&=&0.07\,\textrm{exp}\left(-V/20\right),\\
\beta_{h}(V)&=&1/(\textrm{exp}\left(3-0.1V\right)+1).
\end{eqnarray*}
and for potassium channels,

\begin{eqnarray*}
\alpha_{n}(V)&=&(0.1-0.01V)/(\textrm{exp}\left(1-0.1V\right)-1),\\
\beta_{n}(V)&=&0.125\,\textrm{exp}\left(-V/80\right).
\end{eqnarray*}

In this work we have used for these parameters the standard constant values given in Table (\ref{table1}) \cite{Gerstner2002}.

\begin{table}
\begin{center}
\begin{tabular}{ccc}
\hline
\hline
$x$ & $g_{x}$  $(\textrm{mS}/\textrm{cm}^{2})$ & $E_{x}$ (mV) \\
\hline
$N_{a}$ & 120 & 115 \\
K & 36 & -12 \\
l & 0.3 & 10.6 \\
\hline
\hline \\
\end{tabular}
\end{center}
\caption{The parameters of the Hodgkin-Huxley equations. The membrane capacitance density is $C=1 \, \mu \textrm{F}/\textrm{cm}^{2}$. The voltage scale is shifted so that the resting potential vanishes.}
\label{table1}
\end{table}

The ion currents of sodium, potassium and leakage (mainly chloride) correspond respectively to the three first terms in the right hand of the Eq. (\ref{equ1}), and are generated in response to a change in the respective ion conductances.

\begin{figure}
\begin{center}
\includegraphics[width=1\textwidth]{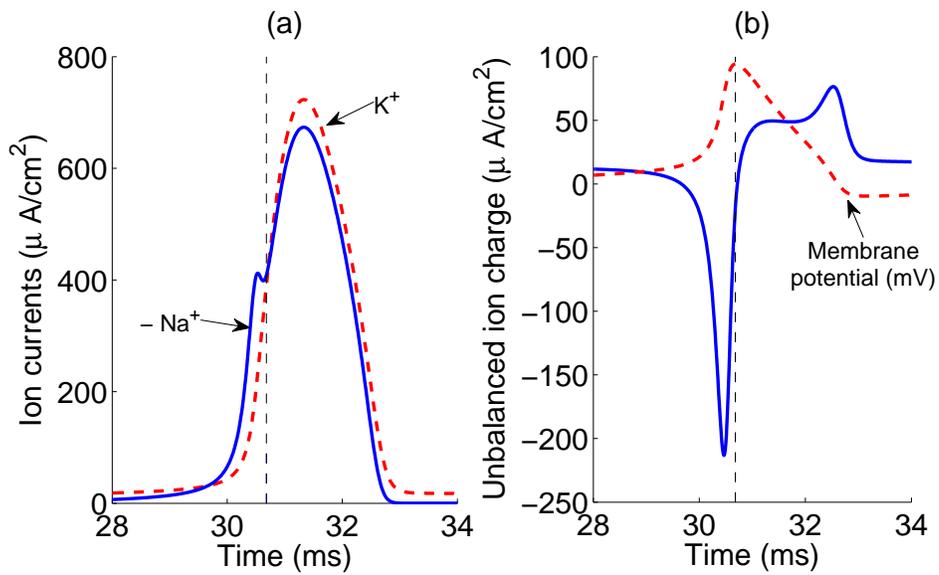}\\
\end{center}
\caption{(a) The currents of sodium and potassium ions during an action potential generated for an external stimulus $I=13 \mu A/cm^2$. (b) The unbalanced ion charge resulting from the overlapping and offsetting of Na$^+$ and K$^+$ flux (the membrane potential is represented in dashed line). The rising and falling phases are separated by the vertical dashed line. The area under the curve in the rising phase region corresponds to the net unbalanced sodium ion charge crossing into the membrane during the rising of the action potential.
} \label{fig1}
\end{figure}

Figure (\ref{fig1}) shows in part (a) the shape of the sodium and potassium currents corresponding to a particular action potential. The sodium current is negative but has been depicted with a positive sign for a better appreciation of the great extent of its overlapping with the potassium current. Note that as sodium and potassium currents are both of positive charges but moving in opposite directions of the cell's membrane they neutralize each other to the extent of their mutual overlap. The sodium charge that is not counterbalanced by simultaneously flowing potassium charge is much smaller for a greater overlap.

The unbalanced current load, represented in Fig. \ref{fig1}(b), consists of two components which occur respectively during the depolarizing and hyperpolarizing phases of the membrane potential action. The integral of the first component of this unbalanced load gives the net Na$^+$ ion charge that is not counterbalanced by simultaneously flowing K$^+$ crossing into the membrane during the rising of the action potential. The integral of the total unbalanced ionic current is directly proportional to the the number of ATP molecules required to restore the resting potential.

For the action potential represented in Fig. \ref{fig1}(b) generated for an external stimulus $I=13 \mu A/cm^2$, the total sodium charge transfer computed as the integral of Na$^+$ current was 1168 nC/cm$^2$  which agree with  the estimate of 1098 nC/cm$^2$ reported recently in \cite{Sengupta2010}. Neutralized currents that account for the overlapping of Sodium and potassium fluxes give rise to an excessive overlap charge of about 1092 nC/cm$^2$. This overlap has been calculated as the difference between the total Na$^+$ current and the depolarizing unbalanced component of Na$^+$ current. As stated in the work of Hodgkin, the squid action potential is very inefficient in the sense that it requires a fourfold Na$^+$ charge compared to the minimum charge necessary to depolarize a pure capacitor \cite{Hodgkin1975}.
The efficiency of sodium entry during the generation of action potential in the squid axon at different temperatures is discussed in Section 3.  The values of sodium and overlap load reported above correspond to a temperature of 6.3 $^\circ$C.

To estimate the energy consumption necessary to restore the resting potential in the Hodgkin-Huxley model, we have used a new approach not based on the ion counting method. Following previous works of finding energy functions of neuron models of chaotic dynamics \cite{Torrealdea2009, Sarasola2005, Sarasola2004}, we have deduced for the model given by Eq. (\ref{equ1}) an energy function representing the analytical expression of the electrochemical energy involved in its dynamics. The procedure followed to find this energy has been reported in detail in \cite{Moujahid2011}, and is summarized below.

It is well known that the Hodgkin-Huxley equation given by Eq.(1) expresses an electrical circuit consisting of capacitor $C$ and three Na, K and L ionic channels, where $g_{N_{a}}$, $g_{K}$ and $g_{l}$ are the maximal conductances, and batteries stand for the Nernst potentials of their corresponding ions. If $V$ is the membrane potential, the total electrical energy accumulated in the circuit at a given moment in time is,

\begin{equation}
H(t) = \frac{1}{2}\, CV^2+H_{N_{a}}+H_{K}+H_{l},
\end{equation}
where the first term in the summation gives the electrical energy accumulated in the capacitor and represents the energy needed to create the membrane potential $V$ of the neuron. The other three terms are the respective energies in the batteries needed to create the concentration jumps in sodium, potassium, and chloride. The electrochemical energy
accumulated in the batteries is unknown. Nevertheless, the rate of electrical energy provided to the circuit by a battery is known to be the electrical current through the battery times its electromotive force.
Thus, the total derivative with respect to time of the above energy will be,
\begin{equation}
\dot{H}(t) = CV\dot{V}+i_{N_{a}}E_{N_{a}}+i_{K}E_{K}+i_{l}E_{l}.
\end{equation}
where $E_{N{a}}$, $E_{K}$ and $E_{l}$ are the Nernst potentials of the sodium, potassium and leakage ions in the resting state of the neuron. And $i_{N_{a}}$, $i_{K}$, and $i_{l}$ are the ion currents of sodium, potassium and leakage, given by,
\begin{equation}
\begin{array}{ll}
i_{N_a}=g_{N{a}}m^{3}h(V-E_{N{a}}),\\
i_{K}=g_{K}n^{4}(V-E_{K}),\\
i_{l}=g_{l}(V-E_{l}),
\end{array}
\end{equation}

If we substitute Eqs. (1) and (4) in Eq. (3), we have for the energy rate in the circuit,
\begin{equation}
\begin{array}{ll}
\dot{H} = VI-g_{N_{a}}m^3h(V-E_{N_{a}})^2-g_{K}n^4(V-E_{K})^2\\-g_{l}(V-E_{l})^2,
\end{array}
\label{equ2}
\end{equation}
which provides the total derivative of the electrochemical energy in the neuron as a function of its state variables. The first term in the right hand summation represents the electrical power given to the neuron via the different junctions reaching the neuron and the other three terms of the summation represent the energy per second consumed by the ion channels.
This equation permits evaluation of the total energy consumed by the neuron and also gives information about the consumption associated to each of the sodium, potassium and leaking channels.

\begin{figure}
\begin{center}
\includegraphics[width=1\textwidth]{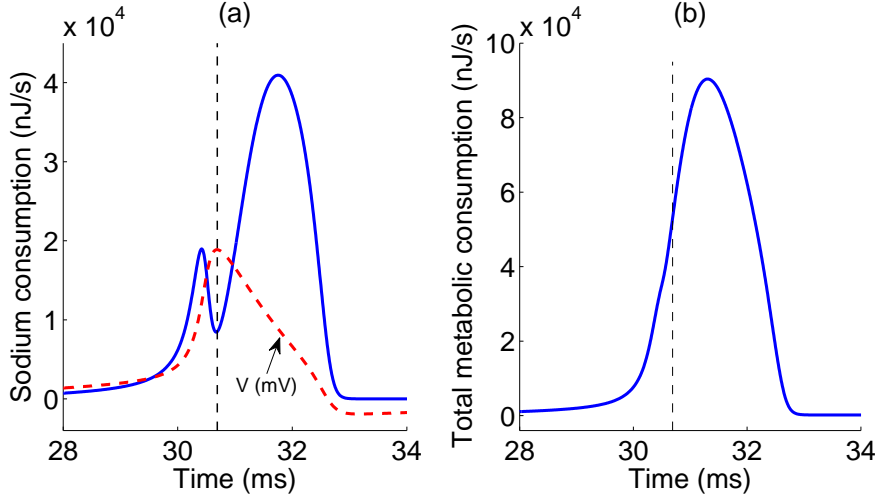}\\
\end{center}
\caption{(a) Electrochemical energy consumption corresponding to sodium ions at an action potential (dashed line) generated for an external stimulus I=13 $\mu A/cm^2$.  Values of membrane potential are scaled with a factor of 200. (b) Total metabolic consumption corresponding to three ion channels at the same action potential. Magnitudes refer to nJ/s. Vertical dashed line in panels (a) and (b) separates between the rising and falling phase regions.}
\label{fig2}
\end{figure}

Figure \ref{fig2}(a) reports the time course of the electrochemical energy consumption in nJ/s corresponding to sodium ion channel for the particular action potential (dashed line) generated at an external constant stimulus I=13 $\mu A/cm^2$. This energy consumption is given by the term $g_{N_{a}}m^3h(V-E_{N_{a}})^2$ (see Eq. 5). The total metabolic consumption given by the sum of the last three terms in Eq.(\ref{equ2}) is reported in Fig. \ref{fig2}(b).
The last three terms of the energy derivative given are negative definite, corresponding to an actual energy dissipation of energy, but has been represented as a positive consumption.

The total consumption of energy per second at the ion channels required to generate one particular action potential must be replenished by the ion pumps and metabolically supplied by hydrolysis of ATP molecules in order to maintain the neuron's activity. The higher demand of metabolic energy associated to the generation of spikes is clearly visible, for instance at $I=13 \mu$ A/cm$^2$ which is the value used to generate the particular action potential analyzed previously, the average of the total metabolic consumption depicted in Fig. \ref{fig2}(b) is about 11.4 $\mu$J/s per membrane unit area.
This consumption must be replenished by metabolic ATP supply.
The number of ATP molecules per membrane unit area hydrolyzed by the Na$^+$/K$^+$ ATPase pump to extrude the Na$^+$  load can be deduced from the amount of Na$^+$ ions crossing the membrane during an action potential, operating with a ratio of 3Na$^+$ per ATP \cite{Attwell2001,Crotty2006}.

The ratio of the total metabolic consumption (in J/s) to one third of the number of Na$^+$ load through the membrane expressed in electronvolts per ATP represents the efficiency of the ATP hydrolysis measured as the free energy provide by the hydrolysis of one molecule of ATP. We will show that our calculation of the actual energy consumption and the number of ATP molecules involved in the generation of an action potential are consistent with relevant data in the literature and that the Hodgkin-Huxley model produces accurate estimates of energy consumption.

\section{Results}
\subsection{Ion currents overlapping}
\label{sec:3}

The overlap of ion currents in the Hodgkin-Huxley model decreases as the temperature increases and
the impact of overlap on the number of ATP molecules required per action potential can
be analyzed rescaling the model equations to include the temperature dependence. In this work, we have adopted the original assumption of Hodgkin and Huxley multiplying the rates of change of the activation $m$, $n$ and inactivation $h$ gating variables by a
factor $k = 3^{(T-6.3)/10}$, $T$ [ $^\circ$C] \cite{Chandler1970}.

\begin{figure}
\begin{center}
\includegraphics[width=1\textwidth]{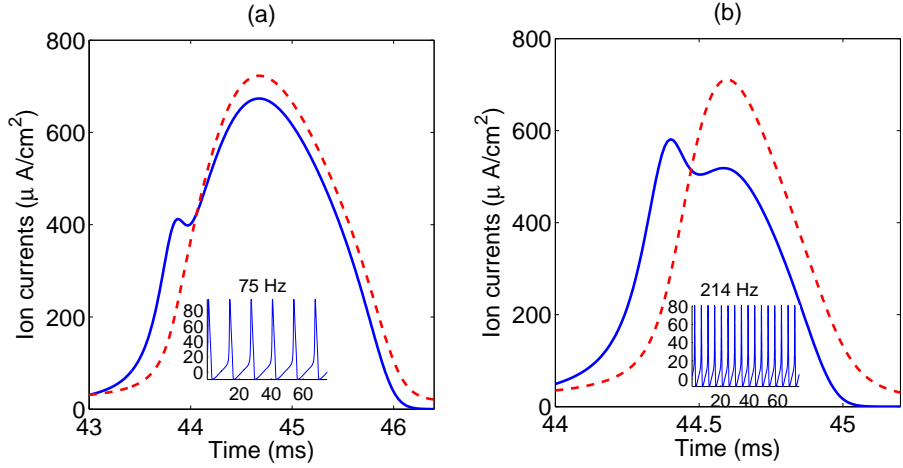}\\
\end{center}
\caption{Simulated sodium (solid line) and potassium (dashed line) currents  at 6.3$^\circ$C (a), and 18.5$^\circ$C (b), showing different degree of overlap. Firing frequencies are 75 Hz and 214 Hz respectively. Sodium current is reversed for comparison.   )
} \label{fig3}
\end{figure}

To illustrate the overlap decrease with increasing temperature, Fig. \ref{fig3} shows instantaneous Na$^+$ and K$^+$ currents elicited by a single spike at respectively 6.3$^\circ$C  and 18.5$^\circ$C.
To perform the simulation we have rescaled the current equations to different temperatures
between 6.3$^\circ$C and 18.5$^\circ$C which is the range of temperatures in the original study
of the squid giant axon by Hodgkin and Huxley. It should be noticed that for higher temperature the firing
regime in the Hodgkin-Huxley model can only be maintained at large values of the injected current.

\begin{figure}
\begin{center}
\includegraphics[width=1\textwidth]{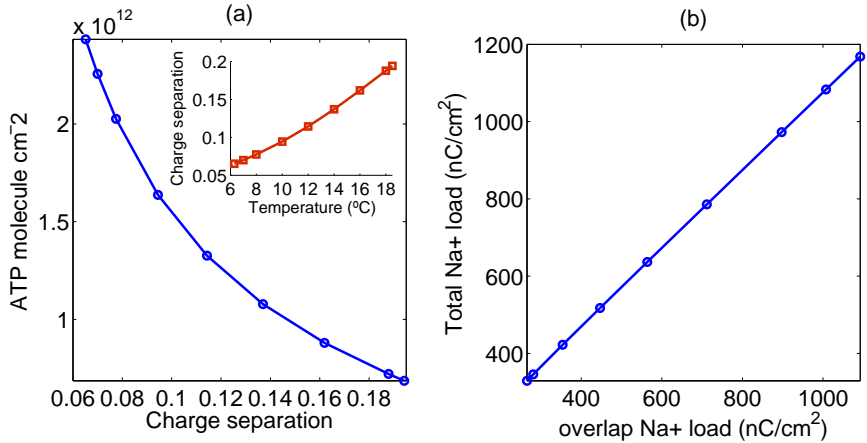}\\
\end{center}
\caption{(a) The number of ATP molecules required per action potential versus charge separation. In the inset charge separation measured as Na$^+$ charge that is not counterbalanced by simultaneously flowing K$^+$ charge divided by total charge per action potential. (b) The relation between the total Na$^+$ load and the overlap charge, calculated as the difference between the total Na$^+$ current and its depolarizing component.
} \label{fig4}
\end{figure}

To quantify the current overlap we have adopted two different measures. Following \cite{Attwell2001}, we calculated the dimensionless charge separation as the Na$^+$ charge that is not counterbalanced by simultaneously flowing K$^+$ charge (depolarizing component of the unbalanced load depicted in Fig. \ref{fig1}(b) divided by total Na$^+$ charge per action potential. The relationship between charge separation and temperature is illustrated in the inset of Fig. \ref{fig4}(a).

As it can be appreciated, charge separation shows a 2.98-fold increase with increasing temperature and varies from 0.0652 at 6.3$^\circ$C to 0.1942 at 18.5$^\circ$C. At this temperature Alle et al. \cite{Alle2009} studying mossy fibres of the rat hippocampus report an average charge separation of 0.769. The respective consumptions per action potential
reflect this different overlap. At 18.5$^\circ$C with an injection current I = 13$\mu A/cm^2$, the Hodgkin-Huxley model of the squid giant axon demands 0.68$\times 10^{12}ATP/cm^2$ to produce one action potential (see Table 2), while according to \cite{Alle2009}, mossy fibers of the rat hippocampus demand only 0.32$\times 10^{12}ATP/cm^2$ per action potential.

The collected results for the number of ATP molecules per unit membrane area to produce one action potential related to charge separation are depicted in Fig. \ref{fig4}(a) . As it can be seen the increase in separation implies a 3.54-fold decrease in the number of ATP molecule/cm$^2$. At 6.3$^\circ$C the Na$^+$ charge transfer per unit membrane area of an action potential in the squid axon is about 1168 nC/cm$^2$, consuming 2.43 10$^{12}$ ATP molecules/cm$^2$. While at 18$^\circ$C, the Na$^+$ load is 346 nC/cm$^2$ consuming 0.72 10$^{12}$ ATP molecules/cm$^2$. So, high frequency firing induced  by high temperature appears to be more efficient in the use of Na$^+$ entry. Table \ref{table2} reports details of values achieved for theses measures at different values of temperature.

The other measure used in this work to quantify the current overlap has dimension of charge and is computed following \cite{Crotty2006} as the difference between the total Na$^+$ load and the depolarizing component of the Na$^+$ load.
Figure \ref{fig4}(b) shows the relationship between the total Na$^+$ load and the overlap load measured in nC/cm$^2$. As it can be seen the total Na$^+$ charge increases linearly with overlap load with a slope close to unity. That is, the overlap load is positively correlated with the total Na$^+$ charge that crosses the membrane resulting in a decisive factor when analyzing the efficiency. The same behavior is observed when considering the relationship between the total unbalanced load computed as the sum of Na$^+$ and K$^+$ currents, and the overlap load.

\begin{figure}
\begin{center}
\includegraphics[width=1\textwidth]{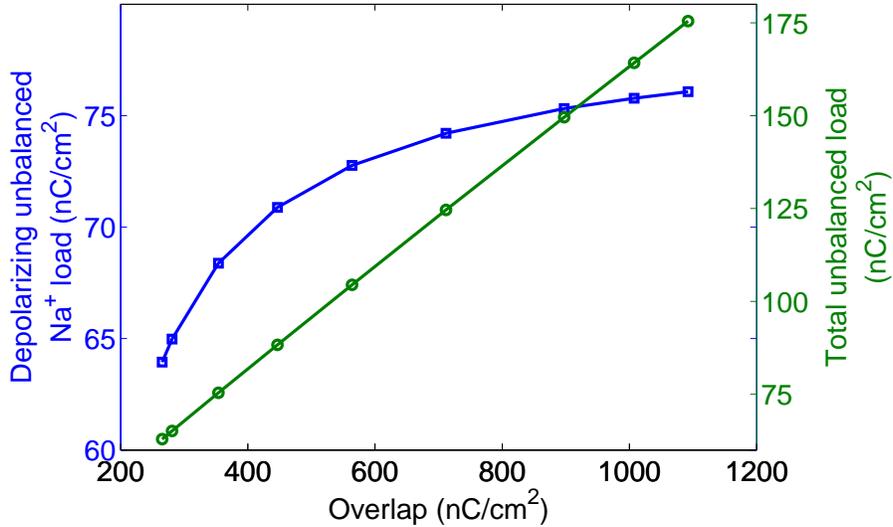}\\
\end{center}
\caption{Hodgkin-Huxley model scaled for different temperatures between 6.3$^\circ C$ and 18.5$^\circ C$. In square markers, the unbalanced Na$^+$ load crossing the membrane during the rising phase of the action potential as a function of overlap load. In circle markers, is represented the total unbalanced load calculated as the sum of sodium and potassium currents related to the overlap load.)
} \label{fig5}
\end{figure}

Figure \ref{fig5} reports, in the right axis, the collected values of the total unbalanced load related to the overlap load. And, in the left axis, the unbalanced component of Na$^+$ load crossing the membrane during the rising phase as a function of overlap. At 6.3$^\circ$C the total unbalanced load is 175 nC/cm$^2$ which is 2.3 times the depolarizing unbalanced Na$^+$ load, while at 18$^\circ$C both measures provide close values around 65 nC/cm$^2$. We observe that for a decrease of overlap load between its maximum value achieved at 6.3$^\circ$C and values around 560 nC/cm$^2$ corresponding to a temperature of about 12$^\circ$C, the depolarized unbalanced load undergoes only a slight decrease. However, further increase of temperature causes the overlap to decrease by 2.12-fold resulting in a 1.13-fold decrease of the depolarizing unbalanced Na$^+$ load.

\subsection{Energy efficiency}

\begin{figure}
\begin{center}
\includegraphics[width=1\textwidth]{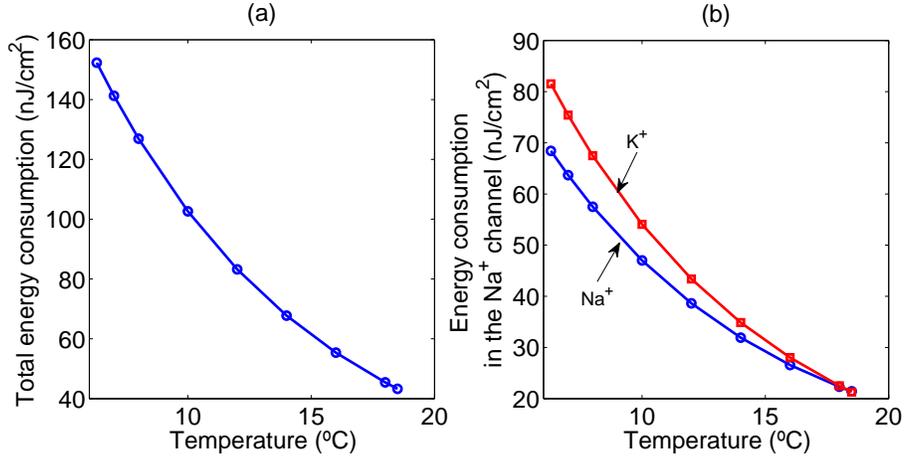}\\
\end{center}
\caption{The Hodgkin-Huxley model scaled for different temperatures between 6.3$^\circ$C and 18.5$^\circ$C. (a) The total metabolic consumption required per action potential versus temperature. (b) The metabolic consumption in the Na$^+$ (circle marker) and K$^+$ (square marker) channels related to temperature. All magnitudes refer to cm$^2$ of membrane.
} \label{fig6}
\end{figure}

The calculated values of the electrochemical energy involved in the dynamics of the Hodkin-Huxley model according to our method are reported in Fig. \ref{fig6}.  Both the total metabolic consumption and the metabolic consumption in the ionic channels show a decreasing pattern as the temperature increases. At 6.3$^\circ$C, the total action potential energy cost is around 152 nJ/cm$^2$, $45\%$ of which is consumed in the sodium channel, and the the rest is mainly consumed in the potassium channel. While at 18$^\circ$C the energy consumed in both channels represents 49$\%$ of the total metabolic consumption. At this temperature the total metabolic consumption experiences a 3.35-fold significant decrease.

The total cost of one pump's cycle, that pumps three Na$^+$ ions out of the cell and two K$^+$ ion in, is computed as the ratio of the total metabolic consumption (given by the last three terms of Eq. 5) to one third of the total number of Na$^+$ load. According to our calculations, the liberated free energy by hydrolyzing one ATP molecule, defined as hydrolysis efficiency, seem to be independent of temperature and shows values around 0.39 eV which is close to other estimates in the literature \cite{Sinkala2006,Nelson2004}.

\begin{figure}
\begin{center}
\includegraphics[width=1\textwidth]{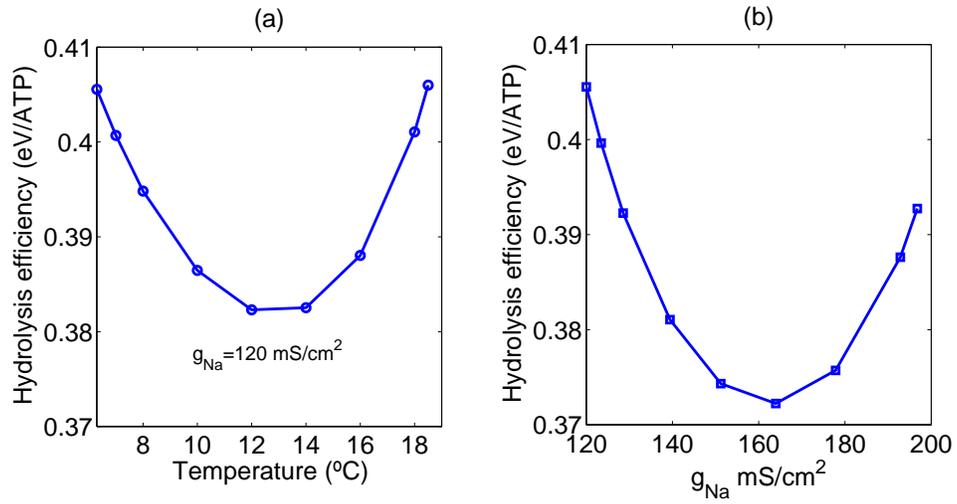}\\
\end{center}
\caption{(a) The ratio of the depolarizing Na$^+$ energy consumption to one third of the depolarizing unbalanced Na$^+$ load as a function of temperature. (b) The ratio of the depolarizing Na$^+$ energy consumption to one third of the depolarizing unbalanced Na$^+$ load as a function of sodium channel conductance. The range of variation of the Na$^+$ channel conductance was obtained multiplying the maximum conductances in the Hodgkin-Huxley model by the factor $k = 1.5^{(T-6.3)/10}$ for temperatures ranging between 6.3$^\circ C$ and 18.5$^\circ C$.
} \label{fig7}
\end{figure}

However, if we consider only the depolarizing metabolic consumption in the  Na$^+$ channel associated to the depolarizing unbalanced Na$^+$ load, the hydrolysis efficiency seem to be affected by temperature showing a parabolic shape with a minimum around 0.38 eV (See Fig. \ref{fig7}(a)). This minimum is observed for a temperature around 13$^\circ$C corresponding to a firing frequency of about 138 Hz.

The same behavior (See Fig. \ref{fig7}(b)) was observed when studying the depolarizing efficiency as a function of the sodium channel conductance. The range of variation of the ion channel conductances was obtained multiplying the maximum conductances $g_{Na}$ , $g_K$ and $g_l$ in the Hodgkin-Huxley model by the expression $k = 1.5^{(T-6.3)/10}$  for temperatures ranging between 6.3$^\circ C$ and 18.5$^\circ C$ according to \cite{Chandler1970}. The maximum depolarizing efficiency occurs for a sodium channel conductance value around 160 mS/cm$^2$  which is close to the biological density conductance. Crotty et al. \cite{Crotty2006}, studying the energetic cost that arise from an action potential using computational models of the squid axon as a function of sodium channel densities, showed that the energy cost associated with the action potential produce a convex curve with a minimum around the same value reported above.

We have seen that the fast-spiking regime in the Hodgkin-Huxley model induced by higher temperatures implies more efficient use of sodium entry, and gives rise to energy efficient action potentials characterized by less overlap between sodium and potassium currents. Since this firing regime could be also induced by raising the external stimulus, it would be interesting to investigate whether the same fast-spiking regime achieved, for one hand, by raising the temperature, and for the other hand, by increasing the external current, contributes in the same way in reducing both the energy consumption and the overlap load.
To do so, we have computed the energy consumption required to generate an action potential and the corresponding firing frequency for different values of the external stimulus I. We have considered values of I ranging between 13 $\mu$A/cm$^2$ and 40 $\mu$A/cm$^2$.

\begin{figure}
\begin{center}
\includegraphics[width=1\textwidth]{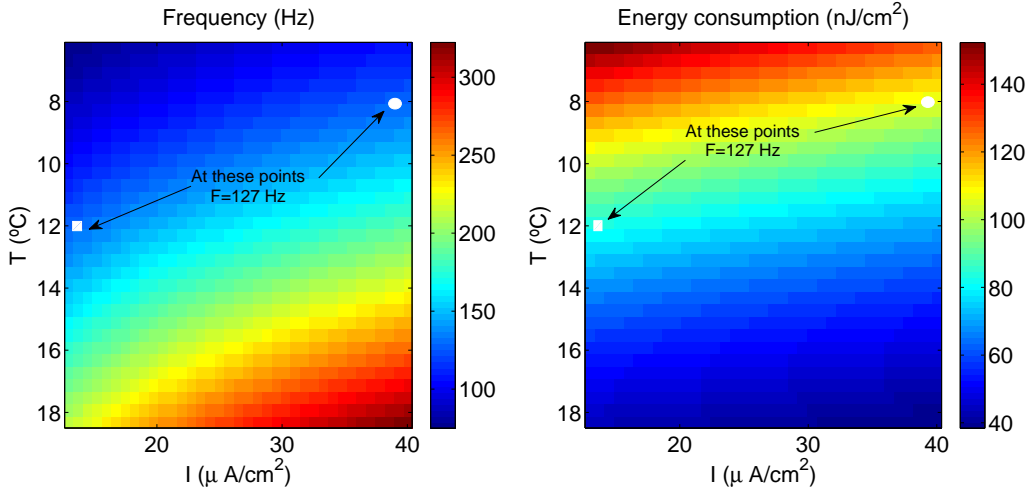}\\
\end{center}
\caption{The squid giant axon. (Left) The spiking regime as a function of temperature and external stimulus. (Right) The total metabolic consumption required per action potential related to temperature and external stimulus. The square and circle markers in the figure correspond to two configurations with the same firing regime.
} \label{fig8}
\end{figure}

As it can be appreciated in Fig. 8 (left panel), the influence of higher stimulus on the firing frequency in the squid giant axon is more significant only for higher temperatures. In fact, at a given temperature, raising the external current causes the firing frequency to increase by 1.46-fold, while at a given external stimulus, the firing frequency experiences a 2.96-fold increase with increasing temperature.
On the other hand, according to our results, the fast-spiking regime induced by higher stimuli is less efficient in generating action potentials expending more energy compared with the same fast-spiking regime induced by higher temperatures. For example, at T=8$^\circ$C and a relatively high external current I=39 $\mu$A/cm$^2$ (see circle markers in Fig. \ref{fig8}), the firing frequency of the squid axon is of about 127 Hz, and are necessary 106.75 nJ/cm$^2$ to generate one action potential. While at a higher temperature T=12$^\circ$C and a low current I=13 $\mu$A/cm$^2$ (see square markers in Fig. \ref{fig8}) corresponding to the same firing regime (i.e, F=127 Hz), the energy consumption is only 83.24 nJ/cm$^2$. This 0.78-fold decrease in energy consumption is accompanied by a 0.76-fold decrease in overlap load between the spike-generating Na$^+$ current and delayed rectifier K$^+$ current. The corresponding values of overlap load are 740.83 nC/cm$^2$ and 563.92 nC/cm$^2$ respectively. This confirms again that the overlap load of voltage-gated currents of Na$^+$ and K$^+$ dominates energy efficiency, and efficient action potentials have little overlap.

\begin{figure}
\begin{center}
\includegraphics[width=1\textwidth]{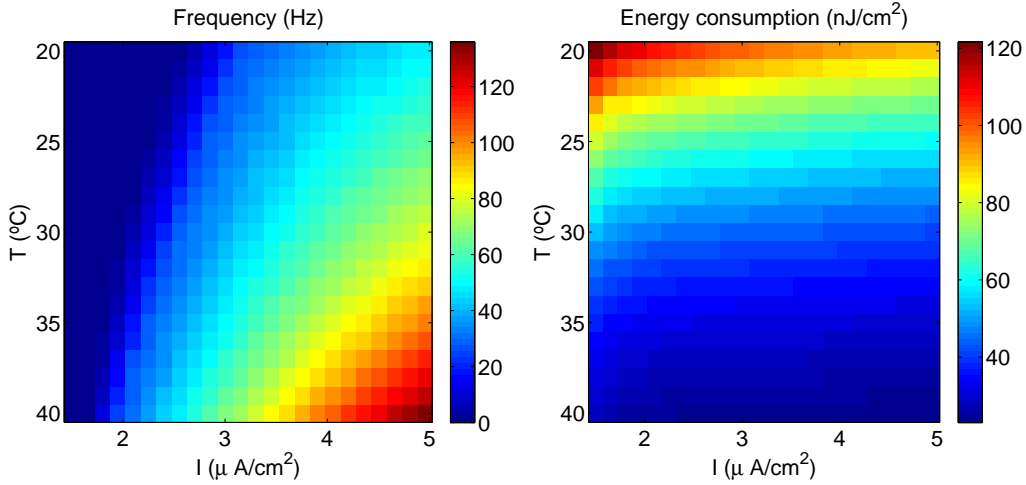}\\
\end{center}
\caption{The regular-spiking model reproducing the typical firing characteristics of regular spiking neurons in ferret visual cortex in vivo. (Left) The firing regime as a function of temperature and external stimulus. (Right) The total metabolic consumption required per action potential related to temperature and external stimulus.
} \label{fig9}
\end{figure}

\section{Discussions}
\label{sec:4}

Our results with simulated spikes of the squid axon show that increased firing frequencies induced by higher temperatures imply more efficient use of sodium entry due mainly to the reduced overlap load between inward Na$^+$ current and outward K$^+$ current. This, corroborates what has been reported recently in \cite{Carter2009,Sengupta2010,Alle2009}, i.e, the most energy efficient action potentials are those generated by Na$^+$ and K$^+$ currents that have substantially reduced overlap.

The values of sodium entry are close to the original values calculated by Hodgkin and Huxley \cite{Hodgkin1952}, and are in nice agreement with values reported recently by Sengupta {\em et al.} \cite{Sengupta2010}. At 6.3$^\circ$C  corresponding to 75 Hz we obtain a sodium influx of 12.12 pmole/cm$^2$ per spike (1168 nC/cm$^2$) while at 18$^\circ$C and 206 Hz the sodium influx is 3.58 pmole/cm$^2$ per spike (346 nC/cm$^2$) which means a 3.38-fold decrease in sodium entry corresponding to a 2.75-fold increase in firing frequency.

Regarding the energy consumption associated to the generation of action potentials in the squid axon, we have found that the hydrolysis of one ATP molecule liberates a free energy with optimum values that range from 0.37 eV achieved for a sodium conductance around 160 mS/cm$^2$ and when only the depolarizing components of both energy consumption and Na$^+$ load are considered. To a value of about 0.39 eV produced when considering the total metabolic consumption associated to the total Na$^+$ load. As stated before, these values of hydrolysis efficiency are in nice agreement with other estimates, which confirms that our method of calculation of the actual energy consumption by the pump and the number of ATP molecules involved are consistent with other data in the literature.

Also, we have found that the fast-spiking regime in the Hodgkin-Huxley model induced by a higher stimulus appears to be less efficient in generating action potentials expending more energy compared with the same fast-spiking regime induced by a higher temperature. Accordingly, the reduction in the overlap of the Na$^+$ and K$^+$ currents is less when the firing frequency is raised by rising the external stimulus. Carter and Bean \cite{Carter2009} have suggested that the primary determinant of differences in Na$^+$ entry efficiency among neurons is their different action potential shapes. Indeed, the shape of the action potential in the squid axon at a given frequency is different depending on whether it has been generated by raising the temperature or the external stimulus.

The findings of this work were validated using others Hodgkin-Huxley-like model neurons, in particular, we have considered the simplest model of regular spiking cells in neocortex which consists of sodium and potassium currents responsible for generating spikes, and an additional slow voltage-dependent potassium current responsible for spike-frequency adaptation. This model generates action potentials which capture the typical firing characteristics of regular spiking neurons in ferret visual cortex in vivo \cite{Pospischil2008}. To carry out the comparison, we performed the same experiment as for the squid giant axon, i.e, rescaling the model equations given in Ref. \cite{Pospischil2008} to include the temperature dependence of the membrane ionic conductances, using values that range between 20$^\circ$C  and 40$^\circ$C which corresponds to the normal range of temperature in these cells. The external current was varied between 1.5 $\mu$A/cm$^2$ to 5 $\mu$A/cm$^2$.
The results (see Fig. 9) show a behavior qualitatively similar to that we observed in the squid axon. i.e, less energy is spent per action potential at higher temperatures than at lower ones due mainly to the reduced overlap between sodium and potassium currents. And, the the efficiency of action potentials is more dependent on temperature than on the external stimulus.
At T=20$^\circ$C and an external high stimulus of about 4.75 $\mu$A/cm$^2$, the regular spiking cell fires at a frequency of about 46.5 Hz, and requires 90.5 nJ/cm$^2$ to generate one action potential. While, for a higher temperature T=36.5$^\circ$C and relatively a small current I=2.75 $\mu$A/cm$^2$, making the cell to fire at the same firing regime (i.e, frequency of about 46.5 Hz), the energy consumption is 29 nJ/cm$^2$ representing only one third of the energy expended when considering higher stimulus. This difference in energy consumption is due to a significant reduction of the overlap load that decreases from 506 nC/cm$^2$ to 88 nC/cm$^2$.

Our principal findings were that the energy consumption required to generate action potentials in the squid giant axon as well as in  the regular-spiking model of cells in neocortex is lower at higher temperatures. Also, we found that for these cells, the fast-spiking regimes induced by higher temperatures are more energy efficient than those induced by higher stimuli.
Finally, we think that the approach considered in this work could bring a new framework to analyze the relationship between energy consumption, temperature and firing frequency in neuronal tissues since it could be potentially used for more relevant models of mammalian brains.

\begin{table*}
\begin{center}
{\small
\begin{tabular}{lcccccccc}
\hline
\hline
Temperature  &   &   &  &   &   &   &   \\
($^\circ$C)  & 6.3 & 8 & 10 & 12 & 14 & 16 & 18 &18.5\\
\hline
Firing rate &   &   &  &   &   &   & &  \\
(Hz)   & 75&   88&  106&  127&  150&  177&  206 &214\\\hline
Metabolic &   &   &  &   &   &   & &  \\
consumption (nJ/cm$^2$) & 152.3&    126.9&    102.6&    83.2 &   67.7 &   55.3 &   45.4 & 43.2\\\hline
Total Na$^+$ &   &   &  &   &   &   & &  \\
load (nC/cm$^2$) & 1168&    973&    786&    637 &   518 &   422 &   346 & 329\\\hline
Overlap$^+$ &   &   &  &   &   &   &  & \\
load (nC/cm$^2$) & 1092&    897&    712&    564 &   447 &   354 &   281 &265\\\hline
Na$^+$  &   &   &  &   &   &   &  & \\
(pmole/spike) & 12.12&    10.09&    8.15&    6.6 &   5.37 &   4.38 &   3.58&3.41 \\\hline
ATP $\times 10^{12}$  &   &   &  &   &   &   & &  \\
(molecule/spike) & 2.43&    2.02&    1.63&    1.32 &   1.07 &   0.87 &   0.72 &0.68\\\hline
\hline
\hline \\

\end{tabular}
}
\end{center}
\caption{Simulation values at different temperatures between 6.3$^\circ$C and 18.5$^\circ$C of the overlap between N$^+$ and K$^+$ ion currents in the squid giant axon. The degree of overlap is measured
as the difference between the total Na$^+$ load and the depolarizing component of the Na$^+$ load per action potential. Their corresponding values of firing frequency, consumption rate, ATP molecules and Na$^+$ entry per spike are also given.
All extensive magnitudes refer to cm$^2$ of membrane.
}
\label{table2}
\end{table*}



\end{document}